\documentclass{article}

\newenvironment{proof}{

\noindent{\bf Proof:}\ }{
}

\newtheorem{theorem}{Theorem}

\newtheorem{lemma}{Lemma}
\newtheorem{corollary}{Corollary}
\newtheorem{definition}{Definition}

\newcommand{\cost}{\mbox{$\mathop{\rm cost}$}}
\newcommand{\size}{\mbox{$\mathop{\rm size}$}}
\newcommand{\credit}{\mbox{$\mathop{\rm credit}$}}

\newcommand{\LL}{\mbox{\sc Landlord}}
\newcommand{\fwf}{\mbox{\sc Fwf}}
\newcommand{\lru}{\mbox{\sc Lru}}
\newcommand{\fifo}{\mbox{\sc Fifo}}
\newcommand{\OPT}{\mbox{\sc Opt}}
\newcommand{\llc}{\mbox{\sc ll}}
\newcommand{\opt}{\mbox{\sc opt}}

\newcommand{\Opt}{\mbox{\sc Opt}}

\begin{document}

\title{\Large On-Line File Caching\thanks{%
    This research partially funded by NSF CAREER award CCR-9720664.
}}
\author{Neal E. Young\thanks{%
    Dartmouth College,
    Hanover NH 03755.
    Akamai Technologies,
    Cambridge MA 02138, USA.
    neal@young.name.
    {\bf \copyright{} 2002, Algorithmica}
    }
  }

\date{}
\maketitle


\begin{abstract}
  Consider the following file caching problem:
  in response to a sequence of requests for files,
  where each file has a specified {\em size} and {\em retrieval cost},
  maintain a cache of files of total size at most some specified $k$
  so as to minimize the total retrieval cost.
  Specifically, when a requested file is not in the cache,
  bring it into the cache and pay the retrieval cost,
  and remove other files from the cache
  so that the total size of files remaining in the cache
  is at most $k$.
  This problem generalizes previous paging and caching
  problems by allowing objects of arbitrary size {\em and} cost,
  both important attributes when caching files
  for world-wide-web browsers, servers, and proxies.

  We give a simple deterministic on-line algorithm
  that generalizes many well-known paging and weighted-caching strategies,
  including least-recently-used, first-in-first-out,
  flush-when-full, and the balance algorithm.
  On any request sequence, the total cost incurred
  by the algorithm is at most $k/(k-h+1)$ times
  the minimum possible using a cache of size $h \le k$.

  For {\em any} algorithm satisfying the latter bound,
  we show it is also the case that
  for {\em most} choices of $k$,
  the retrieval cost is either insignificant
  or at most a {\em constant} (independent of $k$) times the optimum.
  This helps explain why competitive ratios of many on-line paging algorithms
  have been typically observed to be constant in practice.

  \vspace{0.3in}
\end{abstract}

\section{Background and Statement of Results}
The {\em file caching} problem is as follows.
Given a cache with a specified size $k$ (a positive integer)
and a sequence of requests to files,
where each file has a specified {\em size} (a positive integer)
and a specified {\em retrieval cost} (a non-negative number),
maintain files in the cache to satisfy the requests
while minimizing the total retrieval cost.
Specifically, when a requested file is not in the cache,
bring it into the cache, paying the retrieval cost of the file,
and remove other files from the cache
so that the total size of files remaining in the cache
is at most $k$.

Following Sleator and Tarjan \cite{ST-85-2}, we say a file caching algorithm is
{\em $c(h,k)$-competitive} if on any sequence the total retrieval cost incurred
by the algorithm using a cache of size $k$ is at most $c(h,k)$ times the
minimum possible cost using a cache of size $h$.
An algorithm is {\em on-line} if its response to a request
does not depend on later requests in the sequence.

\paragraph{Uniform sizes, uniform costs.~~}
With the restriction that all file sizes and costs
are the same, the problem is called {\em paging}.
Paging has been extensively studied.
In a seminal paper, Sleator and Tarjan \cite{ST-85-2}
showed that least-recently-used
and a number of other deterministic on-line
paging strategies are $\frac{k}{k-h+1}$-competitive.
Sleator and Tarjan also showed that
this performance guarantee is the best possible
for any deterministic on-line algorithm.

A simple randomized paging algorithm called the marking algorithm
was shown to be $2\ln k$-competitive by Fiat et al.~\cite{FiatKLMSY91}.
An optimal $\ln k$-competitive randomized paging algorithm
was given by McGeoch and Sleator \cite{McGeochS91}.
In \cite{Young94}, deterministic paging strategies were
shown to be {\em loosely} $O(\ln k)$-competitive.
This means roughly that for any sequence,
for {\em most} values of $k$,
the fault rate of the algorithm using a cache of size $k$
is either insignificant or the algorithm is $O(\ln k)$-competitive
versus the optimum algorithm using a cache of size $k$.
Similarly, the marking algorithm was shown to be
loosely $(2\ln\ln k + O(1))$-competitive.

\paragraph{Uniform sizes, arbitrary costs.~~}
The special case of file caching when all file sizes are the same
is called {\em weighted caching}.
For weighted caching, Chrobak, Karloff, Payne and Vishwanathan \cite{CKPV-91} 
showed that an algorithm called the ``balance'' algorithm
is $k$-competitive.
Subsequently in \cite{Young94} a generalization of that algorithm
called the ``greedy-dual'' algorithm
was shown to be $\frac{k}{k-h+1}$-competitive.
The greedy-dual algorithm
generalizes many well-known paging and weighted-caching strategies,
including least-recently-used, first-in-first-out, flush-when-full,
and the balance algorithm.

\paragraph{Arbitrary sizes, cost = 1 or cost = size.}
Motivated by the importance of file {\em size} in caching for world-wide-web
applications (see comment below), Irani considered two special cases of file
caching: when the costs are either all equal (the goal is to minimize the {\em
  number} of retrievals), and when each cost equals the file size (the goal is
to minimize the total number of {\em bytes} retrieved).  For these two cases,
Irani \cite{Irani1997} gave $O(\log^2k)$-competitive randomized on-line
algorithms.

\paragraph{Comment: the importance of sizes and costs.}
File caching is important for world-wide-web applications.  For instance, in
browsers and proxy servers remote files are cached locally to avoid remote
retrieval.  In web servers, disk files are cached in fast memory to speed
response time.  As Irani points out (see \cite{Irani1997} and references
therein), file {\em size} is an important consideration; caching policies
adapted from memory management applications that don't take size into account
do not work well in practice.

Allowing arbitrary {\em costs} is likely to be important as well.  In many
cases, the cost (e.g., latency, total transmission time, or network resources
used) will neither be uniform across files nor proportional solely to the size.
For instance, the cost to retrieve a remote file can depend on the {\em
  distance} the file must travel in the network.  Even accounting for distance,
the cost need not be proportional to the size, e.g., because of economies of
scale in routing files through the network.  Further, in some applications it
makes sense to assign different {\em kinds} of costs to different kinds of
files.  For instance, some kinds of documents are displayed by web browsers as
they are received, so that the effective delay for the user is determined more
by the latency than the total transmission time.  Other documents must be fully
transmitted before becoming useful.  Both kinds of files can be present in a
cache.  In all these cases, assigning uniform costs or assigning every file's
cost to be its size is not ideal.\footnote{%
  In many applications the actual cost to access a file may vary with time;
  that issue is not considered here, nor is the issue of cache consistency
  (i.e., if the remote file changes at the source, how does the local cache get
  updated?  The simplest adaptation of the model here would be to assume that a
  changed file is treated as a new file; this would require that the local
  cache strategy learn about the change in some way).  Finally, the focus here
  is on simple {\em local} caching strategies, rather than distributed
  strategies in which servers cooperate to cache pages across a network (see
  e.g.~\cite{KargerLLLLP1997}).}

\newcommand{\tab}{\hspace{0.2in}}
\begin{figure*}
  \begin{center}
  \framebox{\parbox{0.97\textwidth}{
      \setlength{\baselineskip}{1.15\baselineskip}

      \vspace{1ex}

      \underline{Algorithm \LL}

      \vspace{1ex}

      Maintain a real value $\credit[f]$ with each file $f$ in the cache.

      When a file $g$ is requested:

      \setlength{\baselineskip}{.95\baselineskip}

      1. {\bf if} $g$ is not in the cache {\bf then}

      2. \tab {\bf until} there is room for $g$ in the cache:
      
      3. \tab \tab For each file $f$ in the cache,
      decrease $\credit[f]$ by $\Delta\cdot\size[f]$,
      
      4. \tab \tab \tab
      where $\Delta = \min_{f\in\mbox{\footnotesize cache}} \credit[f]/\size[f]$.

      5. \tab \tab Evict from the cache any subset of the files $f$ such that $\credit[f] = 0$.
      
      6. \tab Bring $g$ into the cache and set $\credit[g] \leftarrow \cost(g)$.
      
      7. {\bf else} Reset $\credit[g]$ to any value between its current value
      and $\cost(g)$.
      }}
  \end{center}
  \caption{The on-line file caching algorithm \LL.
    Credit is given to each file when it is requested.
    ``Rent'' is charged to each file in the cache
    in proportion to its size.
    Files are evicted as they run out of credit.
    Step 7 is not necessary for the worst-case analysis,
    but it is likely to be important in practice:
    raising the credit as much as possible in step 7
    generalizes the least-recently-used paging strategy;
    not raising at all generalizes
    the first-in-first-out paging strategy.}
  \label{gdFig1}
\end{figure*}

\paragraph{This paper: arbitrary sizes, arbitrary costs.}
This paper presents a simple deterministic on-line algorithm
called \LL{} (shown in Figure~\ref{gdFig1}).
\LL{} handles the problem of file caching with arbitrary
costs and integer sizes.  The first result is:
\newcommand{\compthm}{
\begin{theorem}
  \label{compthm}
  \LL\ is $\frac{k}{k-h+1}$-competitive for file caching.
\end{theorem}
}
\compthm
This performance guarantee is the best possible for
any deterministic on-line algorithm.\footnote{%
  Manasse, McGeoch, and Sleator \cite{MMS-90} show that
  no deterministic on-line algorithm for the well-known $k$-server problem
  on any metric space of more than $k$ points
  is better than $\frac{k}{k-h+1}$-competitive.
  This implies that, at least for any special case when all sizes are 1
  (i.e.\ weighted caching),
  no deterministic on-line algorithm for
  file caching is better than $\frac{k}{k-h+1}$-competitive.}
File caching is not a special case of the $k$-server problem,
although weighted caching is a special case of both file caching
and the $k$-server problem.

\LL{} is a generalization of the greedy-dual algorithm \cite{Young94}
for weighted caching, which in turn generalizes least-recently-used and
first-in-first-out (paging strategies), as well as the balance algorithm for
weighted caching.  The analysis uses the potential function
$\Phi = (h-1)\sum_{f\in\llc}\credit[f]+ k\sum_{f\in\opt}\cost(f) - \credit[f]$.
The analysis is simpler than that of \cite{Young94}
for the special case of weighted caching.

In an independent work \cite{IC-96}, Cao and Irani
showed that \LL{} (with step 7 raising credit$[g]$ as much as possible)
is $k$-competitive.
They also gave empirical evidence that the algorithm performs well in practice.

\paragraph{This paper: ($\epsilon,\delta)$-loosely $c$-competitiveness.}
In practice it has been observed that on ``typical'' request sequences,
paging algorithms such as least-recently-used, using a cache of size $k$,
incur a cost within a small constant factor (independent of $k$)
times the minimum possible using a cache of size $k$ \cite{Young94}.
This is in contrast to the theoretically optimal competitive ratio of $k$.
A number of refinements of competitive analysis have been proposed
to try to understand the relevant factors.
Borodin, Irani, Raghavan, and Schieber \cite{BorodinIRS95},
in order to model locality of reference,
proposed the {\em access-graph} model
which restricts the request sequences to paths in a given graph
(related papers include \cite{FiatK95,IraniKP1996,FiatR1997}).
Karlin, Phillips, and Raghavan \cite{KarlinPR92}
proposed a variant in which the graph is a Markov chain
(i.e.\ the edges of the graph are assigned probabilities,
and the request sequence corresponds to a random walk)
(see also \cite{LundPR94}).
Koutsoupias and Papadimitriou \cite{KoutsoupiasP94}
proposed the {\em comparative ratio}
(for comparing classes of on-line algorithms)
and the {\em diffuse adversary model}
(in which the adversary chooses a probability distribution,
rather than a sequence, from some restricted class of distributions).

In this paper we introduce a refinement of the aforementioned {\em loosely
  competitive} ratio \cite{Young94} (another previously proposed alternative
model).  The model is motivated by two observations.  First, in practice, if the
retrieval cost is low enough in an {\em absolute} sense, the competitive ratio
is of no concern.  For instance, in paging, if the fault rate
drops much below
$$
\frac{\mbox{time to execute a machine instruction}}%
{\mbox{time to retrieve a page from disk}},$$
then the total
time to handle page faults is less than the time to execute
instructions, so that page faults cease to be the limiting
factor in the execution time.  Similar considerations hold in
other settings such as file caching.  To formalize this, we
introduce a parameter $\epsilon>0$, and say that ``low enough''
for a request sequence $r$ means ``no more than $\epsilon$
times the sum of the retrieval costs'' (the sum being taken
over all requests).  This is tantamount to assuming that
handling a file of cost $\cost(f)$ requires overhead of
$\epsilon\,\cost(f)$ whether it is retrieved or not.

Second, in many circumstances, we do not expect the input sequences to be
adversarially tailored for our particular cache size $k$.  To model this,
rather than somehow restricting the input sequences, we allow all input
sequences but for each, we consider what happens at a {\em typical} cache size
$k$.  Formally, for each sequence, we consider all the values of $k$ in any
range $\{1,2,\ldots,n\}$, and we ask that the competitive ratio be at most some 
constant $c$ for at least $(1-\delta)n$ of these values, where $\delta$ is a
parameter to the model.

Our model, which we dub ``loose competitiveness'', combines both these ideas:
\begin{definition}
  A file caching algorithm $A$ is
  {\em $(\epsilon,\delta,n)$-loosely $c$-competitive}
  if, for any request sequence $r$,
  at least $(1-\delta)n$
  of the values $k\in\{1,2,\ldots,n\}$
  satisfy
\begin{equation}
  \label{loosedef}
  \cost(A, k, r) \le \max\Big\{c\cdot\cost(\OPT, k, r),
  \epsilon\cdot\sum_{f\in r} \cost(f)\Big\}.
\end{equation}
$A$ is {\em $(\epsilon,\delta)$-loosely $c$-competitive}
if $A$ is $(\epsilon,\delta,n)$-loosely $c$-competitive for
all positive integers $n$.
\end{definition}
Here $\cost(A, k, r)$ denotes the cost incurred by algorithm $A$
using a cache of size $k$ on sequence $r$.
\OPT{} denotes the optimal algorithm,
so that $\cost(\OPT, k, r)$ is the minimum possible
cost to handle the sequence $r$ using a cache of size $k$.
The sum on the right ranges over all requests in $r$,
so that if a file is requested more than once,
its cost is counted for each request.

Since the standard competitive ratio grows with $k$,
it is not a-priori clear that any on-line algorithm
can be $(\epsilon,\delta)$-loosely $c$-competitive
for any $c$ that depends only on $\epsilon$ and $\delta$.
Our second result is the following.
\newcommand{\loosethm}{
\begin{theorem}
  \label{loosethm}
  Every $\frac{k}{k-h+1}$-competitive algorithm
  is $(\epsilon,\delta)$-loosely $c$-competitive for
  any $0 < \epsilon,\delta < 1$ and
  $c\,=\, (e/\delta)\ln(e/\epsilon) \,=\, O( (1/\delta)\log(1/\epsilon))$.
\end{theorem}

}
\loosethm
(Throughout the paper $e$ is the base of the natural logarithm.)
The interpretation is that for {\em most} choices of $k$,
the retrieval cost is either insignificant
or the competitive ratio is constant.

This result supports the intuition
that it is meaningful to compare an algorithm
against a ``handicapped'' optimal algorithm
(most competitive analyses consider the case $h=k$).
A strong performance guarantee, even against a handicapped optimal algorithm,
may be as (or more) meaningful than a weak performance guarantee against
a non-handicapped adversary.

Our proof is similar in spirit to the proof in \cite{Young94}
for the special case of paging, but the proof here is simpler,
more general, and gives a stronger result.

Of course the following corollary is immediate:
\begin{corollary}
  \label{loosecor}
  \LL\ is $(\epsilon,\delta)$-loosely $c$-competitive for
  $c \,=\,(e/\delta)\ln(e/\epsilon) \,=\, O( (1/\delta)\log(1/\epsilon))$.
\end{corollary}
This helps explain why the competitive ratios of the many on-line algorithms
that \LL{} generalizes are typically observed to be constant.

For completeness, we also consider randomized algorithms:
\newcommand{\loosethmrand}{
\begin{theorem}
  \label{loosethmrand}
  Let $0\le \epsilon,\delta \le 1$.
  Any $\alpha + \beta\,\ln\frac{k}{k-h+1}$-competitive algorithm
  is $(\epsilon,\delta)$-loosely $c$-competitive for
  $c \,=\, e \alpha+ e\beta \ln [( 1/\delta)\ln(e/\epsilon)]
  \,=\, O(\log [( 1/\delta)\log(1/\epsilon)])$.
\end{theorem}
}\loosethmrand
It is known (e.g.~\cite{youn-91,young:thesis})
that the marking algorithm (a randomized on-line algorithm)
is $(1+2\ln \frac{k}{k-h})$-competitive for paging
and $(1+2\ln k)$-competitive for $h=k$.
It follows by algebra that the marking algorithm is
$1+2\ln 2\,+\,2\ln\frac{k}{k-h+1}$-competitive.
Although a stronger result can probably be shown,
this simple one and Theorem~\ref{loosethmrand}
imply the following corollary:
\begin{corollary}
  The marking algorithm is $(\epsilon,\delta)$-loosely $c$-competitive for
  paging for $c \,=\, e+2e \ln 2 + 2e \ln [(1/\delta)\ln(e/\epsilon)]
  \,=\, O(\log [(1/\delta)\log(1/\epsilon)])$.
\end{corollary}
Finally, we show Theorem~\ref{loosethm} and Corollary~\ref{loosecor}
are tight up to a constant factor:
\newcommand{\tightclaim}{
\begin{theorem}
  \label{tightclaim}
  For any $\epsilon$ and $\delta$ with $0<\epsilon<1$ and $0<\delta<1/2$,
  \LL\ is not $(\epsilon,\delta)$-loosely $c$-competitive
  for
  $c \,=\, (1/8\delta)\log_2(1/2\epsilon)
  \,=\, \Theta((1/\delta)\log(1/\epsilon))$.
\end{theorem}
}\tightclaim

\section{Analysis of \LL.}

\setcounter{theorem}{0}

\compthm
\begin{proof}
  Define potential function
  $$\Phi = (h-1)\cdot\sum_{f\in\llc}\credit[f]
  + k\cdot\sum_{f\in\opt}\cost(f) - \credit[f].$$
  Here $\llc$ denotes the cache of \LL;
  $\opt$ denotes the cache of \OPT.
  For $f\not\in\llc$, by convention $\credit[f]=0$.
  Before the first request of a sequence,
  when both caches are empty, $\Phi$ is zero.
  After all requests have been processed
  (and in fact at all times), $\Phi \ge 0$.
  Below we show that at each request:
  \begin{itemize}
  \item if \OPT{} retrieves a file of cost $c$,
    $\Phi$ increases by at most $kc$;
  \item if \LL{} retrieves a file of cost $c$,
    $\Phi$ decreases by at least $(k-h+1)c$;
  \item at all other times $\Phi$ does not increase.
  \end{itemize}
  These facts imply that the cost incurred by $\LL$
  is bounded by $k/(k-h+1)$ times the cost incurred by $\OPT$.

  The actions affecting $\Phi$
  following each request can be broken down into a sequence of steps, 
  with each step being one of the following.
  We analyze the effect of each step on $\Phi$.
  \begin{itemize}
  \item {\bf \OPT{} evicts a file $f$.}~

    Since $\credit[f] \le \cost(f)$,  $\Phi$ cannot increase.
    
  \item {\bf \OPT{} retrieves a file $g$.}~

    In this step \OPT{} pays the retrieval cost $\cost(g)$.

    Since $\credit[g] \ge 0$,
    $\Phi$ can increase by at most $k\cdot\cost(g)$.

  \item {\bf \LL{} decreases {$\credit[f]$} for all $f\in\llc$.}~

    Since the decrease of a given $\credit[f]$ is $\Delta\,\size(f)$,
    the net decrease in $\Phi$ is $\Delta$ times
    $$(h-1)\,\size(\llc) - k\,\size(\opt\cap\llc),$$
    where $\size(X)$ denotes $\sum_{f\in X}\size(f)$.

    When this step occurs, we can assume that the requested file $g$
    has already been retrieved by $\OPT$ but is not in $\llc$.
    Thus, $\size(\opt\cap\llc) \le h-\size(g)$.

    Further, there is not room for $g$ in $\llc$,
    so that $\size(\llc) \ge k-\size(g)+1$
    (recall that sizes are assumed to be integers).
    Thus the decrease in the potential function is at least $\Delta$ times
    $$(h-1)(k-\size(g)+1) - k(h-\size(g)).$$
    Since $\size(g) \ge 1$ and $k\ge h$,
    this is at least
    $(h-1)(k-1+1) - k(h-1) = 0.$

  \item {\bf \LL{} evicts a file $f$.}~

    \LL{} only evicts $f$ when $\credit[f]=0$.
    Thus, $\Phi$ is unchanged.
    
  \item {\bf \LL{} retrieves the requested file $g$ and sets
      {$\credit[g]$} to $\cost(g)$.}

    In this step \LL{} pays the retrieval cost $\cost(g)$.

    Since $g$ was not previously in the cache (and $\credit[g]$ was zero),
    and because we can assume that $g\in\opt$,
    $\Phi$ decreases by
    $-(h-1)\cost(g) + k\,\cost(g) = (k-h+1)\cost(g)$.

  \item {\bf \LL{} resets {$\credit[g]$} between its current value and $\cost(g)$.}~
    
    Again, we can assume $g\in\opt$.
    If $\credit[g]$ changes, it can only increase.
    In this case, since $(h-1) < k$, $\Phi$ decreases.
\hfill $\diamond$
  \end{itemize}
\end{proof}

\section{Upper Bounds on Loose Competitiveness.}

The following technical lemma is 
at the core of Theorems~\ref{loosethm} and~\ref{loosethmrand}.
\begin{lemma}
  Let $A$ be any $\tau(k,k-h)$-competitive algorithm
  for some function $\tau$ that is increasing w.r.t.\ $k$
  and decreasing with respect to $k-h$.

  For any $b,\epsilon,\delta,n > 0$ ($n$ an integer, $b<\delta n$),
  $A$ is $(\epsilon,\delta,n)$-loosely $c$-competitive for
  \[c \, = \tau(n,b)\,\epsilon^{-(b+1)/(\delta n- b-1)}.\]
\end{lemma}
\begin{proof}
  Fix any request sequence $r$ and $b,\epsilon,\delta, n > 0$.
  Define $c$ as above.
  Say a value $k\in\{1,2,\ldots,n\}$ is {\em bad} if
  \begin{equation}
    \label{bad}
    \cost(A, k, r) > \max\big\{c \cdot\cost(\OPT, k, r),\,
    \epsilon\cdot\textstyle\sum_{f\in r} \cost(f)\big\}.
  \end{equation}
  We will show that at most $\delta n$ values are bad.

  Denote the bad values (in increasing order) $k_0,k_1,\ldots,k_B$.
  The form of the argument is this: on the one hand,
  we show that $\cost(A,k_i,r)$ decreases exponentially with $i$;
  on the other hand, we know that (for each $i$) $\cost(A,k_i,r)$ is
  not too small (e.g.\ smaller than $\epsilon$ times
  $\cost(A,k_0,r)$); together, these will imply that $B$ cannot be too large.

  From the sequence of bad values, select the subsequence
  $k_0,k_{\lceil b\rceil},k_{2 \lceil b\rceil}, \ldots$
  and denote it $k'_0,k'_1,\ldots,k'_{B'}$.
  The properties of this sequence that we
  use are $k'_i - k'_{i-1} \ge b$ for each $i$
  and $B' \ge B / (b+1) $.

  Since $A$ is $\tau(k,k-h)$-competitive,
  choosing $k=k'_{i}$ and $h=k'_{i-1}$ shows that
  \[
  \cost(A, k'_{i}, r) \,\le\, \tau(k'_i,k'_i-k'_{i-1})\, \cost(\OPT, k'_{i-1}, r).
  \]
  From the first term in the maximum in~(\ref{bad}),
  $\cost(A, k'_{i-1}, r) \ge c\cdot\cost(\OPT, k'_{i-1}, r)$.
  The condition on $\tau$ implies
  $\tau(k'_i,k'_i-k'_{i-1}) \le \tau(n,b)$.
  Thus,
  \[
  \cost(A, k'_{i}, r) \,\le\, (\tau(n,b)/c)\, \cost(A, k'_{i-1}, r).
  \]
  Inductively,
  \[
  \cost(A, k'_{B'}, r) \,\le\, (\tau(n,b)/c)^{B'} \cost(A, k'_0, r).
  \]
  That is, for every $b$ bad values, $\cost(A, k_i, r)$
  decreases by a factor of $\tau(n,b)/c$.  
  The rest is algebra.
  As noted before,
  $\cost(A, k'_{B'}, r)> \epsilon\, \cost(A,k'_0, r)$.
  Combining with the above inequality gives \((\tau(n,b)/c)^{B'} > \epsilon,\)
  which (by substituting for $c$ and simplifying) gives
  \[B' < \delta n / (b+1) \,-\,1.\]
  Combining this with $B' \ge B / (b+1) $ gives $B+1 < \delta n$.
  That is, there are fewer than $\delta n$ bad values.
  \hfill$\diamond$
\end{proof}

\loosethm
\begin{proof}
  Fix any $\epsilon,\delta,n > 0$ ($n$ integer).
  We need to show the algorithm is
  $(\epsilon,\delta,n)$-loosely $c$-competitive.
  Let $\tau(k,k-h) = k/(k-h+1)$
  and $b = \delta n/\ln(e/\epsilon) \, - \, 1$.
  If $b \le 0$, then an easy calculation shows $c\ge n$,
  and since the algorithm is $k$-competitive,
  the conclusion holds trivially.

  Otherwise ($b > 0$), we apply the technical lemma.
  With this choice of $b$,
  $\epsilon^{-(b+1)/(\delta n- b-1)} = e$,
  so $c = e\,\tau(n,b)$.
  For this $\tau$ and $b$,
  $\tau(n,b)$ simplifies to
  $(1/\delta)\ln(e/\epsilon)$.
\hfill $\diamond$
\end{proof}

\loosethmrand
\begin{proof}
  Much as in the preceding proof,
  take $\tau(k,k-h) = \alpha+\beta\,\ln(k/(k-h+1))$
  and $b = \delta n/\ln(e/\epsilon) \, - \, 1$.
  If $b \le 0$, then an easy calculation shows
  $c \ge \alpha+\beta\ln n$,
  so the conclusion holds trivially.

  Otherwise ($b > 0$), we apply the technical lemma.
  With this choice of $b$,
  $\epsilon^{-(b+1)/(\delta n- b-1)} = e$,
  so $c = e\,\tau(n,b)$.
  For this $\tau$ and $b$,
  $\tau(n,b)$ simplifies to
  $\alpha+ \beta \ln [( 1 /\delta)\ln(e/\epsilon)]$.
\hfill $\diamond$
\end{proof}

\section{Lower Bound on Loose Competitiveness.}

\newcommand{\pause}{}
\newcommand{\st}{such that~}
\newcommand{\new}{\textrm{new}}

In this section we show the following theorem.

\tightclaim

For the proof
we adapt an unpublished result from \cite{young:thesis}.
We consider the least-recently-used (\lru)
and flush-when-full (\fwf) paging strategies.
(Recall that paging is the special case of file caching when
each size and retrieval cost is 1.)  
We assume the reader is familiar with \fwf\ and \lru,
but just in case here is a brief description of each.
When an item not in the cache is requested
and the cache is full,
\fwf\ empties the cache completely.
In contrast, \lru\ evicts the single item
that was least recently requested.
Figure~\ref{gdFig2} describes how each is 
a special case of \LL.

\begin{figure*}
  \begin{center}
  \framebox{\parbox{0.95\textwidth}{
      \setlength{\baselineskip}{1.15\baselineskip}

      \vspace{1ex}

      \underline{Algorithm \LL\ for the special case of paging}

      \vspace{1ex}

      Maintain a value $\credit[f] \in [0,1]$ with each item $f$ in the cache.

      When an item $g$ is requested:

      \setlength{\baselineskip}{.95\baselineskip}

      1. {\bf if} $g$ is not in the cache {\bf then}

      2. \tab {\bf if} there are no 0-credit items in the cache,

      3. \tab \tab  {\bf then} decrease all credits by the
      minimum credit.

      4. \tab Evict from the cache any subset of the items $f$ such that $\credit[f] = 0$.
      
      5. \tab Bring $g$ into the cache and set $\credit[g] \leftarrow 1$.
      
      6. {\bf else} Reset $\credit[g]$ to any value between its current value
      and $1$.
      }}
  \end{center}
  \caption{\LL\ as it specializes for paging.
    To get \lru, reset $\credit[g]$ to 1 in line 6
    and evict the single least-recently-requested 0-credit
    item in line 4.
    To get \fwf, leave $\credit[g]$ unchanged in line 6
    and evict all 0-credit items in line 4.
    To get \fifo, leave $\credit[g]$ unchanged in line 6
    and evict the single 0-credit item
    that has been in the cache the longest in line 4.
    All of these strategies maintain credits in $\{0, 1\}$.}
  \label{gdFig2}
\end{figure*}

We give the desired lower bound for \fwf.
Since \LL\ generalizes \fwf, the result follows.
This appears unsatisfactory, because it would be natural to restrict \LL\ 
(in line 5) to evict only one file at a time (unlike
\fwf).  However, the same lower bound proof
applies even to a version of \LL\ that has this behavior.
(We discuss this more after the proof.)
Interestingly, the lower bound does {\em not} apply
to \lru.  In fact, for the sequences constructed for the lower
bound, \lru\ is a near-optimal algorithm.

The proof uses the concept of {\em $k$-phases} from the
standard competitive analysis framework.  We define
$k$-phases as follows.  Let $s=s_1 s_2 \ldots s_n$ be any
sequence of requests.  Consider running \fwf{} with a cache of
size $k$ on the sequence, and break the sequence into pieces
(called {\em phases} or {\em $k$-phases}) so that each piece
starts with a request that causes \fwf{} to flush its cache.
Thus, each phase (except the last) contains requests to $k$ distinct items,
and each phase (except the first) starts with a request to an item
not requested in the previous phase.

\smallskip
\noindent{\bf The adversarial sequence.}
Fix any $\epsilon,\delta \ge 0$
with $\epsilon < 1$ and $\delta \le 1/2$.
Define (with foresight)
$c$ as in the theorem
and let $n$ be
some sufficiently large integer.
We will show that \LL\ is not $(\epsilon,\delta,n)$-loosely
competitive.
Define $k_0 = \lceil (1-\delta)n \rceil$.
We will focus on $k$ in the range $k_0,\ldots,n$,
inductively constructing a sequence $s$ such that each
cache size in this range is bad for \fwf\ 
in the sense of Condition~(\ref{bad}).
That is, for each such $k$, we will show
$\cost(\fwf, k, s) > \max\{c\, \cost(\opt, k, s), \epsilon|s|\}$.
The number of $k$'s in the range is $1+n-k_0 > \delta n$,
so this will show the desired result.

In the construction we will build sequences that contain
a special request ``{\bf x}''.  Each occurrence of {\bf x} represents a request
to an item that is not requested anywhere else
(so all occurrences refer to different items).

For the base case of the induction,
we let $s_0$ be a sequence containing $k_0$ special requests {\bf x}.
For the inductive step we do the following.
For $i=0,1,2,\ldots$ let $k_{i+1} = \lceil k_0(1+1/(4c))^i \rceil$
and let $s_{i+1}$ be obtained from $s_i$
by choosing any $k_{i+1} - k_i$ special requests {\bf x}
(including the first one) in $s_i$,
replacing each unchosen {\bf x} with a regular request not occurring elsewhere in $s_i$,
and then appending two copies of the modified string.

For example, if $k_0 = 4$ and $k_1 = 5$,
then $s_0 = \mathbf{x x x x}$
and  $s_1 = \mathbf{x 1 2 3 x 1 2 3}$.

We let the final sequence $s$ be any $s_i$ such that $k_i > n$.
This describes the construction.
The basic useful properties of $s$ are the following:
\begin{lemma}
  (1) Each $s_i$ has length $k_0 2^i$ and references $k_i$ distinct items.

  (2) Any item $r$ introduced in the $i$th inductive step (building
  $s_{i+1}$) has {\em periodicity} $k_0 2^i$ in $s$.
  That is,  for some $j$ with $1 \le j \le k_0 2^i$,
  the positions in $s$ at which $r$ is requested are
  $j, j+k_0 2^i, j + 2\cdot k_0 2^i, j + 3\cdot k_0 2^i, \ldots$.

  (3)
  For each $i$, each length-$k_0 2^i$ contiguous subsequence of $s$
  references $k_i$ distinct items.
\end{lemma}
\begin{proof}
  Properties (1) and (2) above are easy to verify by induction.
  Property (3) follows from properties (1) and (2).
  In particular, in each length-$k_0 2^i$ contiguous subsequence of $s$,
  each item of periodicity $k_0 2^j$ (for $j\le i$)
  is requested $2^{j-i}$ times,
  and each other request is to an item of periodicity larger than
  $k_0 2^i$ that is requested only once in the subsequence.
  Since each length-$k_0 2^i$ contiguous subsequence has this
  structure, each such subsequence references the same number of
  distinct items as the string $s_i$ --- that is, $k_i$ distinct items.
  \hfill$\diamond$
\end{proof}

Using these properties, we show the following: 
\begin{lemma}
  Suppose $n$ is larger than $4c/(1-\delta)$.
  Using any cache size $k$ such that $k_0 \le k \le n$,
  the fault rate of \fwf\ on $s$ is more than $c$ times that of \lru.
\end{lemma}
\begin{proof}
  In the construction of $s_{i+1}$ from $s_i$,
  we were careful to leave the {\em first} special request {\bf x} in $s_i$ alone.
  This ensures that each $k_i$-phase of $s$ is of length $k_0 2^i$
  and starts with a symbol of periodicity greater than $k_0 2^i$.

  From these properties it is easy to calculate the
  fault rates of \fwf\ using a cache of size $k_i$ on $s$.
  The fault rate of \fwf\ is $k_i / (k_0 2^i)$ ---
  each $k_i$-phase has length $k_0 2^i$
  and causes $k_i$ faults.

  The fault rate of \lru\ can be calculated using the following observation.
  \lru\ with a cache of size $k_i$ faults on exactly those items
  of periodicity greater than $k_0 2^i$.
  This is because \lru\ evicts an item $r$ exactly
  when there have been $k_i$ other distinct items requested
  since the last request to $r$,
  and we know (property (3)) that
  between two requests of any item $r$ with periodicity $k_0 2^j$
  there are $k_j-1$ distinct items (other than $r$) requested.

  We can count the frequency of requests to items with
  periodicity greater than $k_0 2^i$ as follows.
  Consider any contiguous subsequence of length $k_0 2^{i+1}$.
  Let $a$ and $b$ be the first and second half of the subsequence,
  respectively (each of $a$ and $b$ has length $k_0 2^i$).
  We know that there are $k_i$ distinct items requested in $a$,
  and $k_{i+1}$ distinct items requested in $ab$.
  But the items requested in $b$ that are not requested in $a$
  are exactly the items of periodicity greater than $k_0 2^i$.
  Thus, there are $k_{i+1}-k_i$ such items in $b$.
  As each is requested exactly once in $b$,
  the frequency of such requests
  (and the fault rate of \lru\ with a cache of size $k_i$)
  is $(k_{i+1}-k_i)/(k_0 2^i)$.

  Thus, for any $i$, using a cache of size $k_i$,
  the ratio of the fault rate of \fwf\ to that of \lru\ is
  \[k_i / (k_{i+1}-k_i).\]
  An easy calculation
  (using the assumption $n > 4c/(1-\delta)$)
  shows this is at least $2c$.

  What about any $k$ such that $k_i \le k \le k_{i+1}$ for some $i$?
  We know that \fwf\ faults $k$ times in each $k$-phase.
  The number of $k$-phases is at least the number of $k_{i+1}$-phases,
  i.e.\ at least $|s|/(k_0 2^{i+1})$.
  Thus, the fault rate is at least $k_i / (k_0 2^{i+1})$ ---
  half the fault rate of $\fwf$ with a cache of size $k_i$.
  For \lru, the fault rate with a cache of size $k$
  is at most the fault rate with a cache of size $k_{i}$.
  Together these facts imply that (for any $k$
  such that $k_i \le k \le k_{i+1}$ for some $i$),
  using a cache of size $k$, the ratio of the fault rate of \fwf\ 
  to that of \lru\ is at least half the ratio when using
  a cache of size $k_i$.
  Thus, the ratio is greater than $c$.
  \hfill$\diamond$
\end{proof}

To finish the proof of Theorem~\ref{tightclaim},
we need to show that the fault rate of \fwf\ remains
above $\epsilon$ for all $k$ such that $k_0 \le k \le n$.
Reasoning as in the previous proof,
the fault rate of $\fwf$ with such a cache size $k$
is at least $k_i / (k_0 2^{i+1})$ for some $i$ where $k_i \le n$.
So we need to show $k_i / (k_0 2^{i+1}) \ge \epsilon$ if $k_i \le n$.
In fact, we show the stronger result that $1 / 2^{i+1} \ge \epsilon$.

The rest is algebra.  In the following we will use the
inequalities $1+x \ge 2^x$ for $x \le 1$ and $1-x \ge 2^{-2x}$
for $x \le 1/2$.

That $k_i \le n$ implies that $i \le 8 \delta c$ by the
following argument.  (Each line follows from the line
before it by the reason given.)
\[
\begin{array}{rcll}
  k_i & \le & n & \mbox{\tab given} \\
  (1-\delta)n(1+1/4c)^i & \le & n &
  \mbox{\tab definition of $k_i$, and $x \le \lceil x \rceil$} \\
  2^{-2\delta} 2^{i/4c} & \le & 1 &
  \mbox{\tab inequalities mentioned above } \\
  i & \le & 8\delta c &
  \mbox{\tab algebra } \\
\end{array}
\]
Using this we will show
$1/ 2^{i+1} \ge \epsilon$,
which implies
$k_i / (k_0 2^{i+1}) \ge \epsilon$.
\[
\begin{array}{rcll}
  8\delta c & \le &\log_2(1/2\epsilon) & 
  \mbox{ definition of $c$} \\
  i &\le& \log_2(1/2\epsilon) &
  \mbox{ $i \le 8 \delta c$ (proven above)} \\
  1 / 2^{i+1} & \ge & \epsilon &
  \mbox{ algebra }
\end{array}
\]
This concludes the proof of Theorem~\ref{tightclaim}.
\hfill$\diamond$
\smallskip

We can modify \fwf\ so that it doesn't evict all items from
the cache at the beginning of the phase, but instead evicts
the 0-credit items (those not yet request this phase)
one at a time but pessimally --- in the order that
they will be next requested.  The modified algorithm only
evicts one page at a time, but, since it still incurs $k$
faults per $k$-phase, the proof of Theorem~\ref{tightclaim}
applies to the modified algorithm as well.  The modified
algorithm is also a special case of \LL.  Thus, the lower
bound applies to \LL\ even if \LL\ is constrained to evict
only as many items as necessary to handle the current request.

\section{Further Directions}

A main open question here seems to be to more tightly
characterize the loose competitiveness of \lru.
A reasonable goal would be to find a non-trivial lower bound
or an upper bound better than the one implied in this paper.
The latter would show that \lru\ is better than \fwf\ in this model.
It would also be nice to characterize the relative loose competitiveness
of \lru\ and  first-in-first-out (\fifo).

Another direction is to find a non-trivial lower bound
for the randomized marking algorithm for paging.
Finally, the lower bounds in this paper apply
to particular on-line algorithms;
what lower bounds can be shown
for {\em arbitrary} deterministic on-line algorithms,
or for {\em arbitrary} randomized on-line algorithms?

\section*{Acknowledgements}
Thanks to Dan Gessel for useful discussions
and to Pei Cao for pointing out to the author
the importance of file size in web caching.

\bibliographystyle{plain}
\bibliography{full,competitive}
\end{document}